# ENERGY AS A PRIMITIVE ONTOLOGY FOR THE PHYSICAL WORLD

J.E. Horvath and B.B. Martins
Universidade de São Paulo USP - IAG Astronomy, Brazil

ABSTRACT: We reanalyze from a modern perspective the bold idea of G. Helm, W. Ostwald, P. Duhem and others that *energy* is the fundamental entity composing the physical world. We start from a broad perspective reminding the search for a fundamental "substance" (perhaps better referred to as *ousía*, the original Greek word) from the pre-Socratics to the important debate between Ostwald and Boltzmann about the energy vs. atoms at the end of the 19th century. While atoms were eventually accepted (even by Ostwald himself), the emergence of Quantum Mechanics and Relativity were crucial to suggest that the dismissal of energy in favor of atoms was perhaps premature, and should be revisited. We discuss how the so-called *primitive ontology* programme can be implemented with energy as the fundamental entity, and why fields (and their quanta, particles) should rather be considered as non-fundamental. We sketch some of the difficulties introduced by the attempt to include gravitation in the general scheme.

1. INTRODUCTION

Philosophers and physicists have always struggled to figure out what the world is made of. Some of the key developments and ideas in this direction are very well-known, starting from the *arché* of Anaximander all the way to the atoms of Dalton and beyond. More than 20 centuries of debate and hard work took a decisive turn around one century ago, when atoms were shown to be real, and seemed to solve the quest for the fundamental nature. However, shortly after this landmark, an accelerated sequence of events prompted the emergence of Quantum Mechanics (QM), a theory that shook the basic accepted notions of the description of the physical world, largely converting the naïve atomic picture, of pre-Socratics and classical physics, into a complex "reality" which challenges us even today (Bohr, 2007)

At the end of the 19th century the atomic theory, after being originally formulated by the pre-Socratics and adopted and reworked by many thinkers, was on the verge of being adopted widely, after the work of John Dalton, that provided an empirical basis for its confirmation. However, the opposition to the existence of atoms was still substantial, and this clash permeated almost every aspect of Classical Physics at that time.

One of the decisive episodes of this debate happened in Germany in 1895. An important gathering of leading scientists at a Lübeck Conference served as a forum for the problem of the fundamental stuff of the physical world. At this Conference, Ludwig Boltzmann and Wilhelm Ostwald clashed over the meaning and features of *energy*, suggested by Ostwald to be a fundamental entity, in opposition to atomic theory. The debate was described as "stiff fight" by one of the main scientists supporting the energy view, George Helm (in fact, Helm preceded Ostwald in the adoption of the energy point of view (Helm, 2000), but apparently his low-profile saved him from Boltzmann's attack). The main point that eventually leaned the participants to the Boltzmann's side (atomic theory) was lately expressed by A. Sommerfeld, present at that time (Sommerfeld, 1944):

> *"It was quite obvious to us that it was impossible to derive the motion equations of a single mass point from an energy equation, to say nothing of optional degrees of freedom."*

Note that at a deeper level, this statement now sounds too classical and much less definitive: we now know that quantum field theory tools to describe a single point mass (very classical at the time, when not even the electron was known) would run into trouble, but to consider "impossible" a macroscopic description of the elementary entities contrasts our modern views. The idea that *fields* (not that popular one century ago) are the adequate language to describe the energy completes the essential of a modern rebuttal of Sommerfeld's straight rejection, as discussed below.

It is very important to stress that the debate between Boltzmann and Ostwald is not only pre-quantum, but also pre-relativistic. Ostwald wanted to discuss the classical energy ontology on the scientific arena, but he was unsuccessful. Nevertheless, George Helm, Pierre Duhem and others supported to various degrees (albeit in a non-apologetically fashion) Ostwald's view (Deltete, 2007), in fact Duhem (Stanford Encyclopedia of Philosophy, 2025), Mach and others had their own versions of the energy role, all them basically opposing the atomic theory. We believe that at the time (and to a large extent, even today), the rooted idea that energy is a *property* of matter led a major fraction of the physicists to its final rejection (see below). This status of a "property" is meaningful in Classical Physics, but not necessarily in Quantum Physics, to where it has been extrapolated. This is not the first time that a confusion induced by classical thinking arose: the same situation happened with the luminiferous ether, an entity supposed to execute the verb "undulate", finally proved to be unnecessary. The wave-particle duality can be thought of another example of how classical thinking sharpens our views: the so-called wave-particle duality may arise from just a human limitation to see beyond the labels attributed to the phenomena (Bunge, 2014). Particles and waves were never in conflict, it was our perception of particular phenomena that attributed a nature we knew from macroscopic objects and made up our perplexity. To straight up all the situation, and formulate/justify the new proposal, we should start by examining the introduction of fields and quanta first.

## 2. QUANTUM THEORY TURN: FIELDS AND QUANTA

The concept that came to confront the materialistic atomism is the *field*. Classical fields were extensively exploited to construct electromagnetic theory, but the real leap was the quantization, achieved much later. Today, the known elementary particles are now understood as *excitations* of the field. Since Special Relativity established that $E = mc^2$, it is immediately clear that their mass is a concentrated form of the energy of the field (in fact, there are many attempts to obtain, for example, electrons as a stable solution of a topological/nonlinear field theory, all them with problems and limited success). But the important lesson is indeed that matter is actually a concentrated and quantized form of energy, a concept very distant from materialistic pre-Socratic/classical view, in which atoms were portrayed as solid cubes, spheres and other forms. We see that a "property" cannot behave in such manner; therefore, a way to recategorize energy was actually needed more than a century ago.

Following the developments of the 20[th] century, the next step was to address *interactions* between elementary particles. At the elementary level, interactions are currently understood as a consequence of the gauge invariance of a Lagrangian $\mathcal{L}$ under an internal symmetry (Ramond, 1990), while intermolecular interactions, etc. are identified as "integrated" versions of elementary interactions with a crescent hierarchy of spatial scales. Note that this picture holds even without quantization, provided the fields are suitably considered (Weinberg, 2021a).

## 3. WHAT IS ENERGY?

### 3.1 Energy in the quantum realm

The whole discussion on the nature and the role of energy at the turn of the 19[th] century was clearly based on the classical notion that energy is a property or even a (pseudo)substance (Sherr et al., 2012). It turned out that soon after the Lübeck debate and the aftermath, a crucial new ingredient changed the view of energy forever. We are referring to the quanta of energy postulated to solve the important "black body" spectral problem (Jeans, 2009). While Max Planck wanted to solve this issue at any cost (Kragh, 2000), he was not ready to accept the quanta as an ultimate reality of the physical world. Einstein took this idea seriously and produced the photoelectric effect results that eventually granted him the Nobel Prize. It is said that Planck was never convinced that his own hypothesis was a form of thinking the whole world: energy comes intrinsically into packages, in multiples of the Planck's constant (or quantum of action) $h$.

We can now attempt a first identification of the energy as a fundamental entity: energy comes in irreducible quanta of the fields, but, instead of energy being a property of the fields, we argue that it can be used to describe the latter, a statement that we shall return to below.

After these developments more than a century ago, another fundamental fact of Quantum Physics, very related to the energy view, was the re-elaboration and refining of the concept of *vacuum*. The quantum vacuum is not "empty", but rather full of *energy* (actually, energy *density*). This is due to the inevitable quantum fluctuations that promote virtual pairs of particles and antiparticles (also unknown at the turn of the 19th century), provided their quantum numbers coincide with the vacuum ones. Matter can "vanish" in the vacuum if it finds a suitable antiparticle, which in turn can be created independently (compare with the ancient pre-Socratic view of Empedocles and others, Hankinson 2001). It follows that Lavoisier was largely right, but what gets transformed is the *energy*, not just matter (because $E = mc^2$, conversions among them are possible). In essence, and without initially knowing the famous mass-energy equality, the irreducible character of the energy was Ostwald's position.

3.2. Are DM and DE just another form of energy?

Let us now briefly address dark matter (DM) and dark energy (DE) (Horvath, 2021): according to their transformation properties, the fundamental features of matter were suggested to be *mass* and *spin*, the two invariants of the Poincaré's group of transformations. Charge was identified as a property, somewhat analogous to what classical philosophy called *primary* and *secondary*, but ultimately we lack of a true deeper comprehension of what a charge really is (Salam, 1979). DM and DE may be components with yet unidentified charges, which may or may not comply with Poincaré's group transformations (unless we need to modify the very spacetime group, for example, because of the existence of a fundamental length, and work within the De Sitter group, Aldrovandi & Pereira, 2009; Horvath, 2021).

In summary, speculating from this point of view into the dark sector, DM and DE with some particular charges could ultimately fall under the category of "energy" as defined above. The problem is restated as "find the charges relevant to hide these unknown components that do not belong to the ordinary SM sector". It is possible, although not popular, that DM and DE do not exist at all, but happen to be an illusion resulting from the projection of extra dimensions (Marteens, 2001) or stemming from a deeper understanding of gravity (Capozzielo & De Laurentis, 2011), but this discussion would take us far from the main scope of this work and will not be addressed.

4. A NEO-ENERGETICS FORMULATION

Since energy and matter are two forms of the same thing, a unified view at the quantum level may be attempted. Classical physicists at the turn of the 19th century thought energy and matter as different entities, and denied a fundamental character to the former. Ostwald inverted this view and was rejected as anti-atomistic, because his Energetics was conceived and seen as an alternative to atomic theory.

As stated, after a century of Quantum Physics, we have plenty of direct evidence of an equivalence between matter and energy, and we see the vacuum energy density as real (Casimir effect, cosmological constant, etc.), proving that particles can pop up from the vacuum as they are excitations (quanta) of fields. This synthesis of matter/energy helps to formulate a *Neo-Energetics* view.

A very crucial first point is that energy is matter (and vice versa), and *not a property* of the latter. It is better understood as a substance-like entity. This unification may also engulf the dark sector as pointed out above, although this last possibility is now speculative at best.

To appreciate the full situation in quantum theory, we know that the energy arises as the expectation value of the Hamiltonian operator $\hat{H}$ for a system in a state labeled as $|n\rangle$, giving the energy of the system as

$$\langle n|\hat{H}|n\rangle = \int_{-\infty}^{\infty} \Psi_n^* \hat{H} \Psi_n d^3x = \langle E \rangle \quad . \tag{1}$$

The Hamiltonian above can be obtained from the basic Lagrangian $\mathcal{L}$ by canonical procedures. For simplicity, we address a scalar field, with a Hamiltonian operator given by

$$\hat{H} = \int d^3x \, [(\dot{\varphi})^2 + (\nabla\varphi)^2 + V(\varphi)] \quad , \tag{2}$$

adding other variables, for example, an electromagnetic field, yields additional terms for the total Hamiltonian. The pure electromagnetic field Hamiltonian is

$$\hat{H} = \int d^3x \left[ (\vec{E})^2 + (\vec{B})^2 \right] \quad . \tag{3}$$

We may think the expressions eqs.(1-3) as defining the fundamental energy of a system $\langle E \rangle$ out from the (auxiliary) variables $\varphi$, $\vec{E}$ and $\vec{B}$, and not as the calculation of a property $\langle E \rangle$ from the same (fundamental) variables. In fact, we see that changing the description would *not* change the energy or its dynamics (Allori, 2021). Hence, we state that energy qualifies as a primitive ontology in the full sense.

Thinking that particles are just quanta of the field, and that they are otherwise *inconsistent* with any relativistic quantum field theory, as stated by Weinberg, Wilczek and other leading theorists of the 20[th] century (Wilczek, 1999, Weinberg, 2021b), it was suggested to dismiss them as a primitive ontology candidate. And while there are proposals to give *fields* that "primitive" status, particles cannot be reduced to fields (see Jaeger, 2023), and a full ontological view would need to be based on both fields and particles. Based on this inconsistency, we suggest that both particles and fields they are at most *bookkeeping devices for the energy* (i.e. non-fundamental variables (Allori, 2013 and 2021]), and not truly fundamental physical entities in the ontological sense.

Energy is shared as determined by the fields carrying charges (quantum properties) associated to the fields, and the fields are a mean to calculate its dynamics, as suggested by Allori (2013). As already stated, within the Standard Model of elementary interactions (Ramond, 1990), *mass* and *spin* (the Poincare group invariants) characterize the fields because they are quantities that should be specified to comply with spacetime invariances (this is essentially Wigner's view of the particle world, Wigner, 1960). The energy is thus elevated from its "property" status in classical physics to a central, fundamental ontological entity in quantum physics. Since quanta are the building blocks of everything, classical objects having energy are automatically accounted (for example, classical potential energy is related to the gravitational charge, whatever it ultimately happens to be). Sommerfeld's classical world is thought to emerge from the quantum level (i.e. Schlosshauer, 2019) through *decoherence,* but not without some caveats being clarified nowadays.

Note again that this view is the reverse of the standard statement "fields are not energy, but they carry energy", which still has an ultimate classical metaphysical character, in the sense of Aristotle. Now we would state "energy is fundamental, fields are a form of handling and tracking the energy". It is entirely conceivable to change the formalism of fields, for example, finding a consistent description in string theory, and the energy content would still be represented by it. However, it is important that the dynamics of the fundamental ontology entity remains invariant, and does not depend on the chosen description (Allori, 2013).

Energy has additional properties that may be considered important for a fundamental ontology status. One of the most important is its overall *conservation*.

At a fundamental level, we know that for each global symmetry there is a conserved charge (Noether's theorem, Baez, 2025), meaning that each field used to track and conduct the energy there is a symmetry that acts as an identifier or label. Charges are properties related to the specific fields. Consider, for example, the Dirac Lagrangian for the electron

$$\mathcal{L} = \bar{\psi}(i\gamma^\mu \partial_\mu - m)\psi \quad , \quad (4)$$

The invariance of the Lagrangian $\mathcal{L}$ when a phase transformation $\psi \to e^{i\alpha}\psi$ is performed (mathematically described by the group of transformations U(1), Ramond 1990) yields a conserved current

$$j^\mu = \bar{\psi}\gamma^\mu\psi \quad . \quad (5)$$

Now if $j^\mu$ is conserved, all its components are. In particular, the time component (zero*th* in this choice of coordinates) is

$$Q = \int d^3x\, \bar{\psi}\gamma^0\psi = \int d^3x\, \psi^\dagger\psi \quad , \quad (6)$$

which is identified with the *electric charge*. Therefore, the electron field is associated with the charge arising from a symmetry property which serves to couple the field to other participants of the electromagnetic interactions (including electromagnetic fields $\vec{E}$ and $\vec{B}$).

In general, we may also state that energy itself is conserved because of the invariance of the system under time translations. This can be checked starting with the operator of time translations $\hat{T}(t) = exp\left(-\frac{i\hat{H}t}{\hbar}\right)$, a function of the Hamiltonian of the system $\hat{H}$. It is quite direct to show that $\hat{T}(t)$ commutes with $\hat{H}$, that is

$$[e^{-i\hat{H}t/\hbar}, \hat{H}] = [\hat{T}(t), \hat{H}] = 0 \qquad . \qquad (7)$$

And since in the Heisenberg picture for any operator $\hat{A}$

$$\frac{d\hat{A}}{dt} = \frac{i}{\hbar}[\hat{H}, \hat{A}] \qquad , \qquad (8)$$

making $\hat{A} = \hat{T}(t)$ proves that $\frac{d\hat{H}}{dt} = 0$ and therefore energy is conserved. In other words, the (whole) energy is the "conserved charge" of the time translation symmetry.

It is relevant to stress here that the very first consistent description of a quantum system, namely Schrödinger's equation, can be viewed as an ultimate realization of *energy conservation* (Griffiths & Schroeter, 2018). In other words, after finding a mathematical entity (the wavefunction $\Psi$), living in an abstract Hilbert space, the dynamics of $\Psi$ is derived by imposing energy to be conserved, yielding the well-known equation

$$\hat{H}\Psi = i\hbar \frac{\partial \Psi}{\partial t} \qquad , \qquad (9)$$

This fact could be considered a rebuttal of Sommerfeld's statement about the lack of adequateness of energy as a starting point for a dynamical description, although at that time nobody would have foreseen such a thing. In this view, the wavefunction is *not* the fundamental entity of the world, it is just a formalism to track the time evolution of the energy (Lewis, 2016). Therefore, we should not worry that the Fock space itself does not square with the manifest image of the physical world (Allori, 2013).

In fact, Allori (2013) has summarized the common elements of scientific explanations repeated here for the sake of definiteness (see also Goldstein, 1998; Dürr, Goldstein & Zanghì, 1992)

> a) Any fundamental physical theory is supposed to account for the world around us (the manifest image), which appears to be constituted by three-dimensional macroscopic objects with definite properties.
>
> b) To accomplish that, the theory will be about a given *primitive ontology*: entities living in three-dimensional space or in space-time. They are the fundamental building blocks of everything else, and their histories through time provide a picture of the world according to the theory (the scientific image).
>
> c) The formalism of the theory contains primitive variables to describe the primitive ontology, and non-primitive variables necessary to

> mathematically implement how the primitive variables will evolve in time.
>
> d) Once these ingredients are provided, all the properties of macroscopic objects of our everyday life follow from a clear explanatory scheme in terms of the primitive ontology.

Against the suggestions of putting the wavefunction as the fundamental entity of a primitive ontology, we stress that the statement that $\Psi$ contains the full information on the state of the system does *not automatically* qualify it as such, and is better seen as a non-primitive quantity in modern terms. We may define $\Psi$ as a "full library bookkeeping device". When a fully relativistic quantum field theory is formulated, and the wavefunction is substituted by field(s), this situation is essentially unchanged. As pointed out in Dürr *et al.* (2004), the role of the wave functions in all these theories is to determine the law of motion for the primitive ontology. Thus, it may be considered to have a law-like, nomological character, and this fact led Dürr *et al.* (2004) to suggest that the wavefunction should be intended as a *physical law*. As it stands, the whole idea of a physical law may suffer a re-elaboration within the latter view, a long-standing task indeed (Lewis, 2016).

We observe that the first of those characteristics highlighted by Allori (2013, 2021) relates in principle the Neo-Energetics proposal with gravitational theories, i.e. the primitive quantity must be connected to the observed reality and be present in the three-dimensional world, rather than relying solely on an abstract and specific mathematical representation (Allori, 2013). If the candidate ontology entity cannot explain observed reality- both microscopic and macroscopic- it is only a mathematical tool useful for a specific domain. Nevertheless, gravitation poses deep problems that we shall sketch below.

5. COMPLICATIONS DUE TO GRAVITATIONAL INTERACTIONS

In the absence of gravitation, the above considerations would ideally suffice to move forward and construct a Neo-Energetics research programme. However, gravitation is a fundamental interaction that shapes the late Universe and all structures within. Therefore, it is necessary to acknowledge that we must address gravitation within the framework of the suggested ontology.

As expected, all the known problems that impeded the treatment of gravitation as a fundamental interaction on the same foot as the rest of elementary interactions appear here. We do not have a consistent quantum gravity theory in spite of many attempts, and a few main approaches can be cited to characterize what has been done and how these constructions relate to the Neo-Energetics proposal.

The "natural" approach in the last century was to consider a quantization of the classical General Relativity, something that could not be made consistent in the same way the other interactions could. This fact led some to postulate that the GR variables are not adequate for quantization, but some new variables could be useful, such as the so-called Ashtekar variables (Ashtekar, 1986). Despite of the advances, this topic cannot be considered as resolved.

On the other hand, a group of researchers choose *not to quantize* gravitation, on the basis that a quantum theory was not possible for it. Conversely, there were attempts to "geometrize" the rest of fundamental interactions, obtaining them as a curvature effect of an extended version of the spacetime. This is indeed the original intention of Kaluza-Klein theories (Overduin & Wesson, 1997 and references therein)

As a third avenue to address this problem, we can count the suggestion of T. Padmadhaban (2010), which states that GR is a kind of emerging theory, akin to fluid mechanics of the gravitational field. Another interesting view has been elaborated by S.L. Adler (1982), in which GR emerges as the long wavelength limit of the true fundamental theory, and possibly related to symmetry breaking in quantum field theory. Many other views of gravity can be quoted, all them bringing difficulties to the quantum realm.

We believe that there is no room to deny that quantum physics rules the microphysical world, and it would be truly remarkable that gravitation could stand alone as a classical theory. But the absence of a consistent theory of quantum gravity precludes the discussion of gravitation on the same foot as the other interactions within the Neo-Energetics framework.

One famous problem with *classical* GR, and many other metric theories indeed, is that there is no well-defined energy-momentum tensor for the gravitation itself. All that can be constructed is a pseudo-tensor, which is invariant under a restricted set of transformations only. Einstein (1918), Landau and Lifshitz (1975) and others have constructed versions of this quantity, but did not obtain a true tensor that can be meaningful for any coordinate system. We believe that, taken the Neo-Energetics seriously, this supports the idea of gravitation as an effective theory. A true fundamental theory should possess a well-defined tensor $T_{\mu\nu}^{grav}$ associated to the gravitation, and in addition a quantization of the theory should be possible as it was with the other interactions.

Thinking of what is available, it is clear that General Relativity works very well, but that does not mean that it is valid for all domains. Looking forward, as several attempts have been made to develop a Unified Theory (Weinberg,

2011), gravitation and the quantum world areas remain disjoint. The existence of different forms of energy cannot be described entirely by General Relativity or any other classical gravitation theory. In other words, GR description has its limits in terms of explanation, as it applies only to a subset of the whole reality. But ideally, starting from energy, it should be possible to explain the quantum world, the large-scale Universe and all smaller scales down to the microscopic realm in between. From this point of view, the conservation of four-momentum energy in the General Relativity theory may not be a true problem: at first, it may seem like a statement of non conservation of energy, but, because it is a just a mathematical tool specific to a domain, this cannot be considered a physical fact, at least until a Unified theory of a better version of gravitation valid in the quantum world can be constructed and a proper form of $T_{\mu\nu}^{grav}$ defined.

In some sense, Quantum Field theories are likely closer to the primitive ontology programme as long as they are directly related to energy (and its conservation) and three fundamental interactions - electromagnetic, weak, and strong interactions– have already been described by them. There are expectations that gravity will also be explained by these theories, with the discovery of a consistent way of quantization. With a full definition of energy, we can possibly derive all things that exist, at least in principle.

6. SUMMARY AND CONCLUSIONS

A summary of this work is then

- Energetics emerged at the "wrong time", and was mainly regarded as an alternative to atomism. Immediately after atoms were accepted, the focus shifted because quantum theory emerged to deal with them. Materialistic atomism was deeply challenged by QM, and ultimately "particles" were shown to be incompatible (in their original sense) with any relativistic formulation (Wilczek, 1999; Weinberg, 2021).

- Fields were elevated by many to the status of fundamental entities in the 20[th] century, while particles were understood as just quanta of the fields. However, a direct correspondence of the field with energy, and the existence of alternatives to the dynamics suggest that they can be considered as a kind of bookkeeping description (point c) above). In the language of primitive ontology, fields are a non-primitive element (Allori, 2013; 2021)

- Energy can be reinstalled as the fundamental entity of the physical Universe (point a) above) now that quanta and deeper insights like Noether's theorem are available. Energy is "the stuff the world is made of" (*ousía*) and fields (labeled by their charge), and a mean to write down the energy dynamics (point b) above). This may be extended to the so-called "dark sector" having hidden charges (to be confirmed).

- Finally, atoms can be (laboriously) constructed out of energy quanta, represented by fields (point d) above), which means that in principle the classical view of the ontology should be reversed: energy composes objects, it is not a property of them.

- Gravitation is not yet incorporated into the quantum domain, and its problems for doing this are well-known, but unsolved.

This is the sketchy minimal conceptual framework of a Neo-Energetics related to the original Ostwald suggestion, but heavily reprocessed and reformulated by quantum theory along one century, and still challenged by gravitation. A long path has been travelled to clean up the original objections to Energetics, and its 21th century quantum version may help understand the physical world in a totally new fashion.

As a final remark we stress that, although our aim was to discuss the role of energy as a primordial entity, there is a connection to be made and understood with its exchange, in particular with irreversible processes (see Kondepudi & Prigogine, 2014). Irreversible Thermodynamics deals with the latter using energy as a basic material. Therefore, a unified view of Energy and its transformations from the most fundamental point of view as discussed above could and should be attempted.

foton@iag.usp.br
martins.bianca@usp.br